\documentstyle[12pt,aaspp4]{article}
\newcommand{\ea}{{et~al.}}
\newcommand{\IUE}{{\it IUE}}
\newcommand{\dmod}{$(m - M)_{0}$}

\newcommand{\logl}{$\log L$}

\newcommand{\lsun}{L$_{\sun}$}
\newcommand{\msun}{M$_{\sun}$}
\newcommand{\muv}{$m_{162}$}

\newcommand{\teff}{T$_{\rm eff}$}

\begin{document}  

\title{Ultraviolet Imagery of NGC 6752: A Test of Extreme Horizontal Branch 
Models}

\author{Wayne B. Landsman\altaffilmark{1},
Allen V. Sweigart\altaffilmark{2}, 
Ralph C. Bohlin\altaffilmark{3}, 
Susan G. Neff\altaffilmark{2},
Robert W. O'Connell\altaffilmark{4}, 
Morton S. Roberts\altaffilmark{5},
Andrew M. Smith\altaffilmark{2}, and Theodore P. Stecher\altaffilmark{2}}

\altaffiltext{1}{Hughes STX Corporation, NASA Goddard Space Flight Center,
Laboratory for Astronomy and Solar Physics, Code 681 Greenbelt, MD 20771}

\altaffiltext{2}{NASA Goddard Space Flight Center,
Laboratory for Astronomy and Solar Physics, Code 681 Greenbelt, MD 20771}

\altaffiltext{3}{Space Telescope Science Institute, 3700 San Martin Drive,
 Baltimore, MD 21218}

\altaffiltext{4}{University of Virginia, Department of Astronomy, 
P.O. Box 3818, Charlottesville, VA 22903}

\altaffiltext{5}{National Radio Astronomy Observatory, 520 Edgemont Rd.,
Charlottesville, VA 22903}
\vspace{0.1in}

\begin{abstract}

   We present a 1620 \AA\ image of the nearby globular cluster NGC 6752
obtained with the Ultraviolet Imaging Telescope (UIT) during the Astro-2
mission of the Space Shuttle {\em Endeavour} in 1995 March.
An ultraviolet-visible color-magnitude diagram (CMD) is derived for 216 stars
matched with the visible photometry of Buonanno et al.\ (1986).    This 
CMD provides a nearly complete census of the hot horizontal branch (HB)
population with 
good temperature and luminosity discrimination for comparison
with theoretical tracks.

The observed data show good agreement with the theoretical zero-age horizontal
branch (ZAHB) of \markcite{sweig96} Sweigart (1996) for an assumed reddening 
of E(B--V) = 0.05 
and a distance modulus of 13.05.   The observed HB luminosity width is in
excellent agreement with the theoretical models and supports the single star
scenario for the origin of extreme horizontal branch (EHB) stars.
However, only four stars can be identified as post-EHB stars, 
whereas almost three times this many are expected from the HB number counts.
If this effect is not 
a statistical anomaly, then some non-canonical effect may be decreasing
the post-EHB lifetime. 
The recent non-canonical models of Sweigart (1996), which have helium-enriched
envelopes due to mixing along the red giant branch, cannot explain the 
deficit of post-EHB stars, but might be better able to explain their 
luminosity distribution.

\end{abstract}


\section{Introduction}                                                       
 
The evolution of extreme horizontal branch (EHB) stars
has attracted considerable interest in recent years due to their importance
for understanding the hot stellar population in both metal-poor and metal-rich
stellar systems (e.g.\ \cite{cal89}, \cite{cast94}).   According to 
canonical theory, EHB stars retain only a thin  ($< 0.02$ \msun)
inert hydrogen envelope due to prior mass loss on the red giant branch (RGB), 
and spend their core-helium burning lifetime at high effective temperatures
(20,000 K $\leq$ \teff $\leq$ 35,000 K).
Following the exhaustion of their central helium, EHB stars 
do not return to the asymptotic giant branch (AGB), but instead follow
AGB-manqu\'e tracks, spending the rest 
of their pre-white dwarf lifetime at high temperatures and luminosities.
Because of their sustained hot phases,
EHB and post-EHB stars are the most promising candidates for producing the 
ultraviolet excess observed in 
elliptical galaxies (\cite{greg90}, \markcite{dorm95}Dorman, O'Connell, \& 
Rood 1995, \cite{yi95}).

Several scenarios have been suggested to explain the origin of the EHB stars.
In the single star scenario, EHB stars are produced by extensive mass loss
along the RGB, which reduces the envelope mass to the very small values 
required by canonical EHB models.   This scenario predicts that the EHB 
should appear as a well-defined extension of the blue horizontal branch (BHB)
in the color magnitude diagram (CMD).   The principal 
difficulty with this scenario is
the fine tuning of the mass loss process needed to produce the narrow range of 
EHB envelope masses.   To avoid this fine-tuning problem, 
\markcite{dcruz96} D'Cruz et al.\ (1996) have suggested that some EHB stars 
might be ``hot He-flashers'', i.e. stars which evolve off the RGB to high
effective temperatures before igniting helium.   
Such stars would settle onto a
blue hook at the hot end of the HB.    Several binary 
scenarios have also been proposed, involving either Roche 
lobe overflow along the RGB or the merger of a double helium white dwarf system 
(see \cite{bail95} for a review).
These scenarios predict a wide range in the EHB mass and therefore a wide range
in luminosity.

Theoretical EHB models have not been 
well-tested even in the globular clusters.     The main observational 
difficulty has been the derivation of accurate luminosities and temperatures for
a large sample of EHB and post-EHB stars,
for comparison with theoretical predictions.
In a standard (V, B--V) CMD, the EHB is observed as a 
nearly vertical blue tail, and the effect of an increase in temperature is
nearly degenerate with that of a decrease in luminosity.    Ultraviolet
observations can provide the needed temperature discrimination, but, while
previous ultraviolet experiments have yielded some intriguing discrepancies with
canonical hot HB models (\markcite{whit94}Whitney et al.\ 1994, 
\markcite{dixon96}Dixon et al.\ 1996),
they have lacked either the  sensitivity 
or the spatial resolution for a clean confrontation with the theory.

The recent non-canonical HB models of \markcite{sweig96}Sweigart (1996) 
provide a further impetus for an observational test of EHB and post-EHB 
evolution.
These models include the dredge-up of helium from the hydrogen shell on the RGB, as 
possibly suggested by the observed abundance anomalies in globular cluster red
giants (e.g.\ \markcite{lang95}Langer \& Hoffman 1995).
A star which undergoes such helium mixing will arrive on the HB with an 
enhanced envelope helium abundance and therefore will lie blueward of its
canonical location.
In this scenario, the high effective temperatures of the EHB stars are the
due to a high envelope helium abundance, which considerably increases the
envelope masses.
Thus the problem of fine-tuning the mass loss process is avoided. 

NGC 6752 is a nearby (\dmod\ = 13.05), lightly reddened ( E(B--V)  $\sim$ 0.04)
intermediate metallicity ([Fe/H] = --1.64) globular cluster with a blue HB 
and a large population of EHB stars
(\markcite{buon86}Buonanno et al.\ 1986, hereafter B86).   
The quoted distance for NGC 6752 is taken from the recent HST observations
of the white dwarf cooling curve by 
\markcite{renz96}Renzini et al.\ (1996, hereafter R96), and thus is 
independent of any assumptions concerning the HB luminosity.
In 1995 March,  we imaged NGC 6752 at 1620 \AA\ 
using the Ultraviolet Imaging Telescope (UIT), which was part of the 
Astro-2 Space Shuttle payload.   The UIT field of view ($40'$) is
well matched to the cluster size, and its solar-blind CsI detectors suppress 
the dominant cool star population and allow hot HB stars to be detected into
the center of the cluster.
In this {\em Letter},  we combine the UIT photometry with the visible 
photometry of B86 to derive temperatures and luminosities for a 
large sample of EHB and post-EHB stars for comparison with theoretical 
predictions.

\section{Observations}

The UIT uses image-intensified film detectors and has a 
spatial resolution of about $3''$ (\cite{stech92}). 
NGC 6752 was observed with the B5 filter, which has a 
central wavelength of 1620 \AA, and a width of about 225 \AA\ (\cite{stech92}).
Only the deepest (781s) image (FUV2619) obtained on 08 Mar 1995 is used here
(Figure 1, Plate Lxx).
The magnitudes, \muv, given in this {\em Letter} are defined as
$-2.5 \times \log f_{\lambda} - 21.1$, where $f_{\lambda}$ is the mean flux
through the B5 filter.    The UV photometry was derived by PSF fitting
using an IDL implementation of the DAOPHOT algorithms of 
\markcite{stet87}Stetson (1987), 
with the error analysis modified for use with film.     The complete UV
photometry catalog, along with a discussion of the radial distribution of the
hot HB stars, will be reported in a subsequent paper.

The UIT image of NGC 6752
shows five stars significantly brighter than the other 354 detected 
hot stars.    Three of these stars
were classified as post-EHB stars by 
\markcite{moeh96}Moehler, Heber, \& Rupprecht (1996), while one is 
interior to the annulus studied by
B86, and is designated here as UIT-1.
Only one of these stars (B1754) had been 
previously observed with \IUE, and so we have obtained
short-wavelength (SWP), low-dispersion IUE spectra of the other
four (Table 1).    The \IUE\ spectra each show a hot continuum 
but have insufficient S/N to allow certain identification of any spectral 
features.
The UIT photometry is used to derive \teff\ and \logl\  in Table 1, because
the \IUE\ spectra of three of the 
post-EHB candidates (B1754, B2004, UIT-1) are contaminated by neighboring hot 
HB stars.

The absolute calibration was determined by comparison with the fourteen 
HB stars observed with \IUE\ by \markcite{cacc95} Cacciari et al.\ (1995) 
plus the four new ones
reported here.    The \IUE\ spectra were processed using the NEWSIPS reduction
(\cite{nichols96}), but with the absolute calibration corrected to the 
\markcite{bohl96}Bohlin (1996) scale.
There is a standard deviation of 0.075 
mag between the \IUE\ and UIT fluxes, which is about what
is expected for the two instruments.     

Temperatures are derived from the \muv\ -- V color using the LTE model
atmospheres of \markcite{kur93} Kurucz (1993) 
for [Fe/H] = --1.5 and assuming E(B--V) = 0.05.    This reddening value was
adopted to yield the best agreement between our \teff\ values and those 
derived from Balmer line fitting by \markcite{moeh96} Moehler et al.\ (1996),
and is close to the value of E(B--V) = 0.04 tabulated by 
\markcite{pete93}Peterson (1993).   
There is evidence for an increasing discrepancy between the two methods of 
determining \teff\ for \teff\ $>$ 30,000 K (with Balmer line fitting yielding
higher temperatures), perhaps because NLTE effects 
are more important at these high temperatures, or perhaps because 
the V magnitudes are less reliable for the hottest (and faintest) HB stars.

There is a 97\% overlap between the hot (B--V $< 0.22$) stars in the B86 
catalog, and the stars detected on the UIT image in the same area of the
cluster (roughly between $1.5'$ and $9.5'$ from the cluster center).
Ultraviolet fluxes could not be determined for three HB stars 
on the 
UIT image that are too close to the 
heavily saturated foreground star HD 177999.
The spatial resolution of UIT
was insufficient to obtain useful photometry for the hot HB stars B1525 and 
B1532, which are separated by $2.8''$.
Finally, two stars detected on the UIT image 
could not be matched with any blue stars in B86,
and instead appear to be matched with ``red straggler'' stars with 
composite colors.  B4840 (V = 13.64, B--V = 0.87) is likely a composite of a 
hot HB star and a red giant, while B1370 (V = 16.09, B--V = 0.33) is likely
a composite of a hot HB star and a subgiant.
Further observations are needed to determine whether these composite colors
indicate a physical association.  

\section{Discussion}

Figure 2 shows an ultraviolet-visible CMD for the 216 UIT stars in common 
with B86 together with the canonical zero-age HB (ZAHB) from
\markcite{sweig96}Sweigart (1996) for a 
scaled-solar metallicity of Z $= 5 \times 10^{-4}$ and a main-sequence helium
abundance of Y = 0.23.   
As noted by B86, the CMD shows a relative deficit of HB stars near 
\muv\ -- V = --2.7 ($\log$ \teff\ $\sim$ 4.25). 
The two cool stars (B1624 and B2044) which fall considerably below the
ZAHB are probably either non-members or extreme blue stragglers.
The distribution of hot stars shows excellent 
agreement with the theoretical ZAHB for the assumed distance and reddening, 
except that the hottest stars appear to be too faint by $\sim$ 0.1 -- 0.2 mag.
This offset could be explained if the core mass in the EHB stars was $\sim$
0.01 -- 0.02 \msun\ smaller than the canonical value.    
In any case, Figure 2 suggests that the difference in the ZAHB luminosity 
between the EHB and BHB may be somewhat larger than predicted by canonical models.
We note that 
the hot He flasher stars discussed by \markcite{\dcruz96}D'Cruz et al.\ (1996)
are predicted to lie up to $\sim$ 0.1 mag below the canonical ZAHB.
However, such stars are constrained  to a small temperature range around 
$\log$ \teff\ $\sim$ 4.5 (\muv -- V $\sim -4$), and thus should appear 
as a blue hook at the hot end of the EHB.   No such feature is evident in 
Figure 2.

Our observational data are plotted in the theoretical plane in Figure 3
together with selected evolutionary 
tracks.
Figure 3 shows 
that the predicted luminosity width of the HB is in excellent agreement 
with the observations, with the width increasing from about 0.1 dex 
near $\log$ \teff\ = 4.0 to about 0.25 dex near
$\log$ \teff\ = 4.5.    The EHB in NGC 6752 thus appears to be an extension
of the BHB, as predicted by the single star scenario for the origin of the EHB
stars.

We next consider the four post-EHB stars\footnote{The star B2485 
with $\log$ \teff\ $\sim$ 4.2 and $\log$ L $\sim$ 1.8
in Figure 3 is a possible fifth post-EHB star, although it appears
to be somewhat too cool and faint to arise from post-EHB tracks.}
in Figure 3.
The evolutionary tracks with 
ZAHB effective temperatures greater than $\log$ \teff\ $\sim$ 4.30 
show a long-lived, hot, post-EHB phase.   The theoretical lifetime of this 
post-EHB phase is typically 0.15 -- 0.20 of the EHB lifetime itself. 
Since
the CMD contains 63 stars hotter than $\log$ \teff\ = 4.30, one would therefore 
expect about ten post-EHB stars, compared to the four actually found.
We have done Monte Carlo simulations to determine the probability of 
finding only four post-EHB stars.    
Stars are assigned a random age along the nearest evolutionary track,
starting with a ZAHB distribution matched to the observed number and 
temperature distribution of the HB stars in NGC 6752.
In a set of 1000 such simulations, we find a median number of 11 post-EHB
stars, and find fewer than five post-EHB stars in only 2.2\% of the 
trials.

The deficit of post-EHB stars in NGC 6752 also appears to persist for the
stars for which we do not have optical photometry and hence colors. 
There are 134 sources on the UIT image outside of the region studied by B86,
including 103 sources in the core, and 31 sources on the periphery  of the
cluster.    But among these 134 sources, only UIT-1 is sufficiently bright
to be a good post-EHB candidate.  
Note that a deficit of post-EHB stars in the core might be at least 
partially explained by the deficit of EHB stars reported there by
\markcite{shar95} Shara et al.\ (1995).

If the deficit of post-EHB stars in NGC 6752 is not a statistical anomaly,
what changes might be required in the HB models?   
One possibility
would be to decrease the post-EHB lifetime relative to that of the 
EHB phase.    
It has sometimes been suggested that this lifetime ratio could be decreased 
by EHB models which include ``breathing pulses'' 
(e.g.\ \cite{cal89}, \cite{renz88}), which are sudden increases 
in the size of the convective 
core near the end of the HB evolution.   Whether or not breathing pulses actually
occur, however, remains an open question (\cite{dr93}).
Moreover, \markcite{sweig94}Sweigart (1994) has found that
breathing pulses do not necessarily 
reduce the post-EHB lifetime below its canonical
value.   

Alternatively, the deficit of post-EHB stars might be explained by an 
increase in the critical temperature at which an HB star follows an 
AGB-manqu\'e track, instead of evolving rapidly back toward the AGB.    This 
explanation is supported by the high effective temperatures of the 
four post-EHB stars, suggesting that they are descendants of only the 
hottest ($\log$ \teff $>$ 4.5) HB stars.     
However, again it is not 
apparent what changes in the canonical models would produce such a change in the
post-HB morphology.   
One ingredient missing from the canonical models is
radiation-driven mass loss, which  has been suggested as the cause
of the change in surface abundance as stars evolve from helium-poor sdB stars 
on the EHB, to helium-rich sdO stars in the post-EHB phase (\cite{mac94}).
However, radiation-driven mass loss  would 
remove additional mass from the envelope and hence would {\em increase} the 
number of stars that evolve along AGB-manqu\'e tracks.

We next compare the present results with the helium-mixed models of 
Sweigart (1996).
Figure 4 shows two representative
EHB and post-EHB tracks: one for a canonical sequence (left panel) and one for a
helium-mixed sequence (right panel).   
As discussed below, the
helium-mixed sequence is shown  using a best-fit distance modulus of 
\dmod\ = 13.30, rather than the value of \dmod\ = 13.05 determined by R96.
The canonical sequence in Figure 4 predicts a luminosity gap of $\sim 0.3$ dex 
between the end of the HB and the beginning of the post-EHB phase.   This gap
is a consequence of the interior readjustment that occurs as an HB star 
exhausts its central helium fuel and begins helium burning in a shell.  
The surface luminosity during the post-EHB phase gradually increases 
as the helium shell moves outward in mass, resulting in a rather large
predicted luminosity range ($\sim 0.6$ dex) for the post-EHB stars.

The helium-mixed sequence in Figure 4 
has a high
envelope helium abundance ($Y_{env} \approx 0.5$) due to dredge-up of helium
during the preceding RGB phase, and consequently a much larger 
envelope mass than in canonical models of similar effective temperature 
($\sim 0.03$ \msun versus $\sim 0.006$ \msun).   
Due to this larger envelope 
mass the hydrogen shell reignites at the end of 
the HB phase.   This, together with the increase in the helium-burning 
luminosity at that time, produces both a larger luminosity gap 
of $\sim 0.5$ dex between the HB and the post-EHB phases, and a smaller
luminosity range for the post-EHB stars. 
Figure 4 shows that the four post-EHB stars do cluster
near the luminosities predicted by the helium-mixed track, 
but the observational and statistical uncertainties are large.

As noted above, the helium-mixed ZAHB is more luminous than the canonical
ZAHB, and thus  requires a larger distance modulus 
to fit the NGC 6752 data.
However, the ZAHB luminosity of a helium-mixed EHB sequence depends on the 
mass of its helium core, which, in turn, depends on some extent on the
mixing algorithm used along the RGB.   The helium-mixed sequence in Figure 4,
for example, would be consistent with 
the R96 distance scale if its core mass of 0.501 \msun\ were 
smaller by $\sim$ 0.02 \msun.   It is entirely possible that different mixing
algorithms might produce such changes in the core mass.

Additional observations could help determine whether the deficit of post-EHB
stars in NGC 6752 is a statistical anomaly, and if any modifications are 
needed to the canonical HB tracks.    
We have recently obtained a wide-field UBV CCD mosaic of NGC 6752 
with the goal of improving  upon the photographic photometry of B86.
In addition, HST photometry of the cluster core (e.g. \cite{shar95}) 
should allow the CMD to be extended to the entire cluster.

\acknowledgments

We are pleased to acknowledge the many people involved with the 
Astro-2 mission who made these observations possible.    RWO
acknowledges NASA support through grants NAG5-700 and NAGW-4106 to the
University of Virginia.   AVS acknowledges NASA support through grant 
NAG5-3028. We thank S. Moehler for showing us her work prior to publication, 
and B. Dorman and V. Dixon for useful comments.    We also thank Dr. Yoji 
Kondo and the staff of the \IUE\ observatory for their assistance with the 
\IUE\ observations.

\clearpage

\begin{figure}
%\plotone{figure1.eps}
\caption{(Plate xx) UIT 1620 \AA\ image of NGC 6752.    North is up and East
is to the left.  The heavily saturated source $4'$ southwest of the cluster 
is the foreground  star HD 177999 (V=7.4, B9 II-III).   The 216 stars used in
the CMD are marked with boxes.   (xxx-admin: available as a separate gif file)}

\end{figure}

\begin{figure}
\plotone{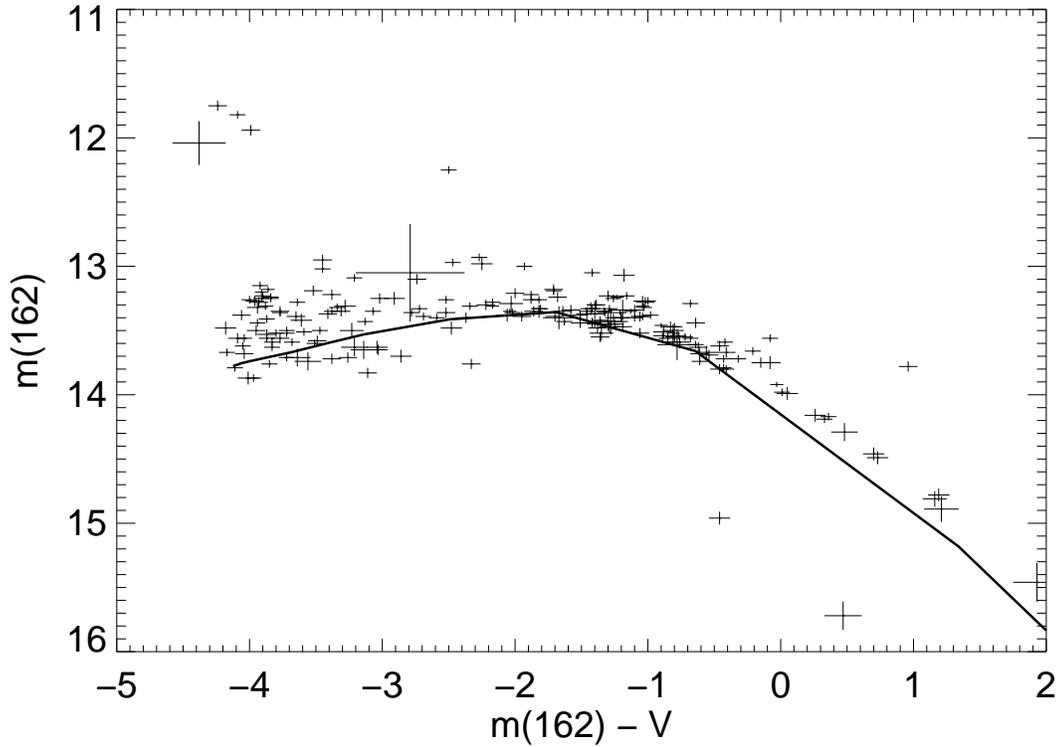}
\caption{An ultraviolet-visible CMD for 216 stars in NGC 6752.    The error
bars are derived from the ultraviolet photometry, and assume a  0.03 mag
uncertainty in the visible photometry.
The solid line shows the canonical ZAHB of Sweigart (1996) for [Fe/H] = --1.5,
assuming \dmod\ = 13.05 and E(B--V) = 0.05.}

\end{figure}

\begin{figure}
\plotone{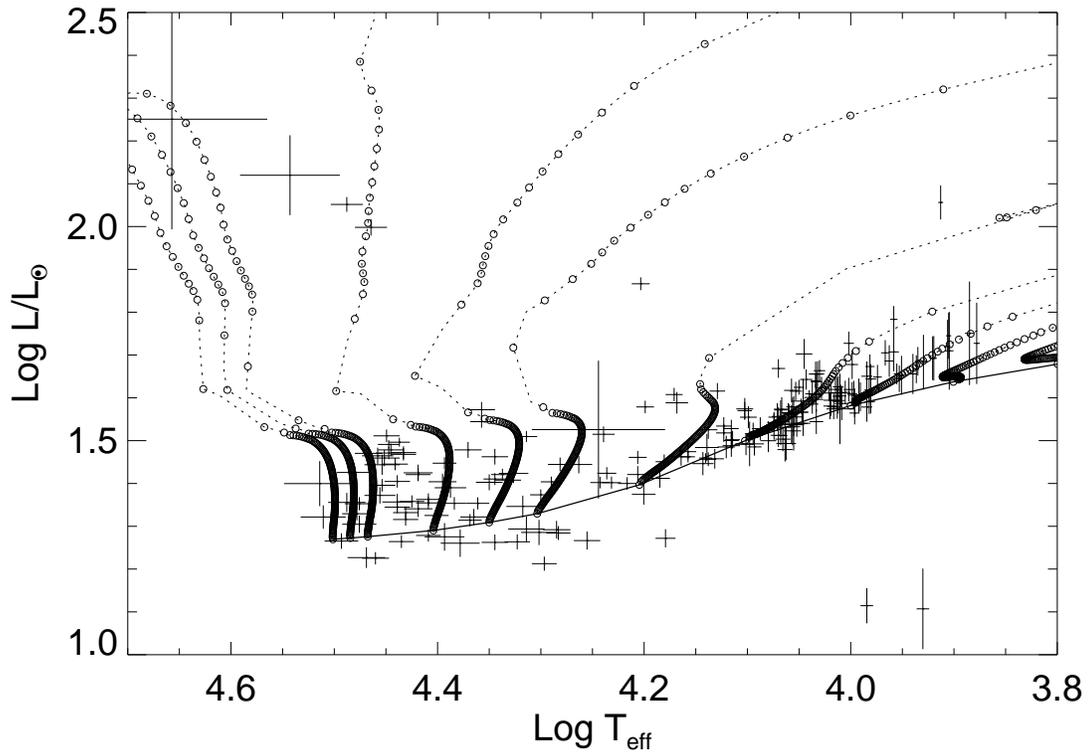}
\caption{The UIT data from Figure 2 is shown transformed to the 
theoretical plane for \dmod\ = 13.05 and E(B--V) = 0.05. 
The error bars are derived from the photometric errors propagated into the  
($\log$ \teff, $\log$ L) plane.
Also shown is the theoretical ZAHB of Sweigart (1996) along with 
selected evolutionary tracks.
The rate of evolution along the tracks is indicated by
the small circles, which are separated by a time interval of $10^6$ yr.}

\end{figure}

\begin{figure}
\plotone{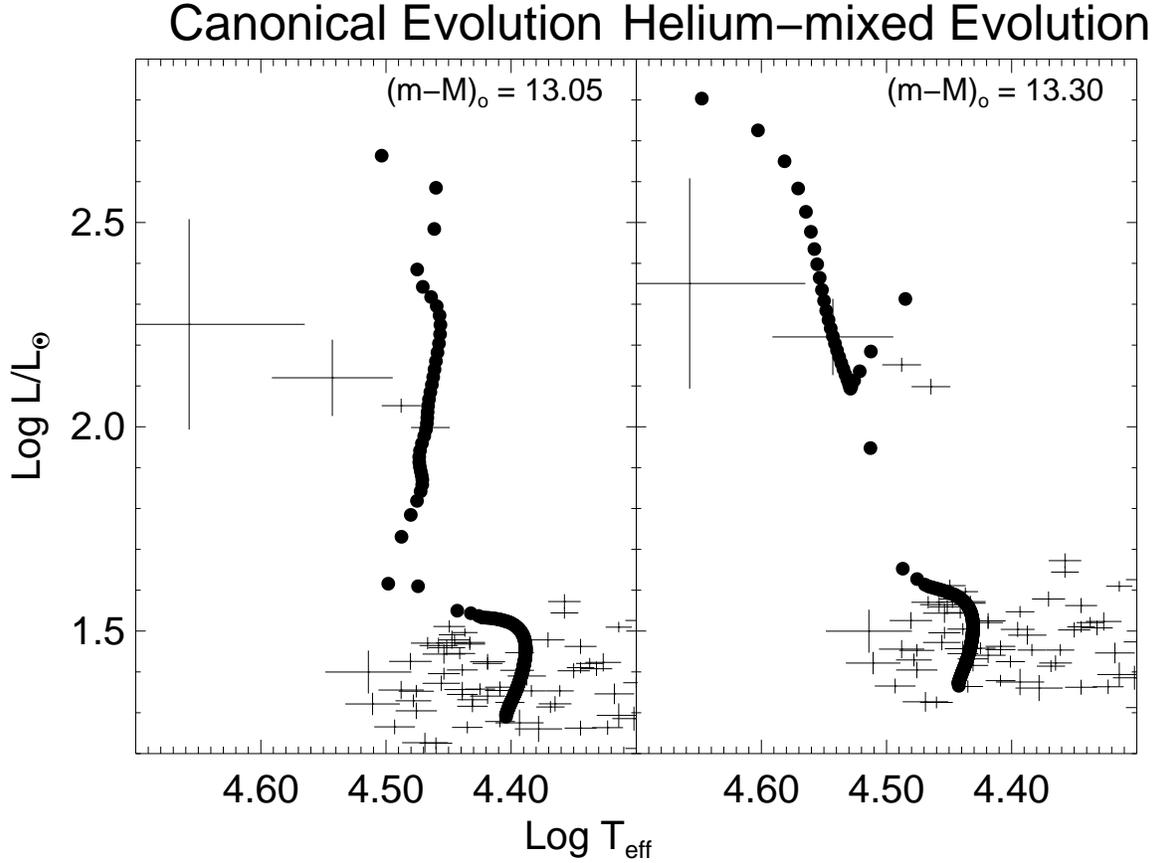}
\caption{The evolution of an EHB star is shown for a 
canonical HB model (left) and a helium-mixed model of Sweigart (1996) 
with Y$_{env} \approx$0.5 (right).   The rate of evolution is marked every 
$5 \times 10^5$ yr.
The error bars show the UIT photometry transformed to the theoretical plane
assuming \dmod\ = 13.05 for comparsion with the canonical model, and
\dmod\ = 13.30 for comparison with the helium-mixed model.}

\end{figure}

\clearpage

\tightenlines

\begin{deluxetable}{rrrrllll}  
\tablecaption{Post-EHB Candidates in NGC 6752}
\tablehead{
 \colhead{Name}  & \colhead{SWP} &  
\colhead{V}  & 
\colhead{B--V} & \colhead{\muv} & \colhead{$\log$ \teff} & \colhead{$\log$ L/\lsun} }
\startdata
 B852\tablenotemark{a} & 55397 & 15.91 & --0.28  & 11.89 & 4.49 & 2.05 \\
 B1754\tablenotemark{a} & 19441 & 15.99  & --0.24  & 11.82 & 4.54 & 2.12 \\
 B2004\tablenotemark{a,b} & 55384 & 16.42 & --0.31 & 12.11:  & 4.66: & 2.25: \\
 B4380 & 55548 & 15.93 & --0.14 & 12.01 & 4.46 & 2.00 \\
 UIT-1\tablenotemark{c} & 55383 &       &       & 11.42  &  &  \\ 

\enddata

\tablenotetext{a}{sdO optical spectrum, \markcite{Moeh96}Moehler et al.\ 1996
and Moehler (personal communication)}
\tablenotetext{b}{uncertain UIT and \IUE\ fluxes due to contamination from the 
 nearby ($2.5''$ distant) hot HB star B1995}
\tablenotetext{c}{RA (2000): 19 10 54.01 Dec (2000) --59 59 46.2}

\end{deluxetable}

\end{document}